# Revisiting first-principles thermodynamics by quasiharmonic approach: Application to study thermal expansion of additively-manufactured Inconel 625

Shun-Li Shang,[1,2,*] Rushi Gong,[1] Michael C. Gao,[2] Darren C. Pagan,[1] and Zi-Kui Liu[1]

1. Department of Materials Science and Engineering, The Pennsylvania State University, University Park, PA 16802, USA
2. Computational Science Engineering Directorate, National Energy Technology Laboratory, Albany, OR 97321, USA

*E-mail: sus26@psu.edu (S. L. Shang).

Contact Dr. Shang for Supplementary Excel file if need.

**Abstract**

An innovative method is developed for accurate determination of thermodynamic properties as a function of temperature by revisiting the density functional theory (DFT) based quasiharmonic approach (QHA). The present methodology individually evaluates the contributions from static total energy, phonon, and thermal electron to free energy for increased efficiency and accuracy. The Akaike information criterion with a correction (AICc) is used to select models and model parameters for fitting each contribution as a function of volume. Using the additively manufactured Inconel alloy 625 (IN625) as an example, predicted temperature-dependent linear coefficient of thermal expansion (CTE) agrees well with dilatometer measurements and values in the literature. Sensitivity and uncertainty are also analyzed for the predicted IN625 CTE due to different structural configurations used by DFT, and hence different equilibrium properties determined.

Advances in density functional theory (DFT) based first-principles calculations enable the accurate prediction of thermodynamic properties without the fitting of phenomenological models [1]. For example, the DFT-based quasiharmonic approach (QHA) has been widely used to predict Helmholtz energy $F(V,T)$ as a function of volume ($V$) and temperature ($T$) for a given configuration [2,3],

$$F(V,T) = E(V) + F_{vib}(V,T) + F_{el}(V,T).$$  Eq. 1

where $E(V)$ is the static total energy at 0 K without the zero-point vibrational energy, $F_{vib}(V,T)$ the vibrational contribution, and $F_{el}(V,T)$ the thermal electronic contribution [2,3]. Beyond thermodynamic properties for a given configuration calculated with DFT, the multiscale entropy approach (recently termed the zentropy theory [4,5]) can accurately predict experimentally observed macroscopic anharmonicity and emergent properties such as critical points, negative thermal expansions, and Curie temperatures, through integrating quantum mechanics (i.e., Eq. 1) and statistical mechanics (via the partition function), where the key input for zentropy is QHA-based free energies for both the ground-state and the non-ground-state configurations [4,5]. For accelerated development of new alloys for advanced manufacturing (including additive manufacturing), rapid prediction of properties for screening and larger length-scale models is needed, and the QHA is an attractive option. Here, we present a new method for evaluating thermodynamic properties using QHA with improved accuracy. This method is applied to evaluating the linear coefficient of thermal expansion (CTE) of the additively manufactured nickel-based superalloy Inconel 625 (IN625).

In practice, the following two DFT-based QHA schemes have been frequently used to evaluate Helmholtz energy, i.e., $F(V,T)$, and in turn, equilibrium properties (see details below Eq. 2) at each $T$ under a given external pressure $P$ ($P = -\frac{\partial F}{\partial V} = 0$ GPa used herein and hence Helmholtz energy equals to Gibbs energy).

- **QHA Scheme 1 (Sch. 1)**: All contributions of $E$, $F_{vib}$, and $F_{el}$ are predicted by DFT at the same given volumes. Then, the $F(V,T)$ curve is fit at each $T$ after summing these contributions at each volume. This scheme requires successful DFT calculations for all contributions at all given volumes, and the predicted curves are usually not smooth.
- **QHA Scheme 2 (Sch. 2)**: Each individual contribution to Helmholtz energy is predicted,

then each contribution is fitted separately in terms of DFT data points at the same or different volumes, i.e., $E$ at 0 K only, while $F_{vib}$ and $F_{el}$ at each $T$. At each temperature, the $F(V,T)$ curve is then a summation of the three fitted curves.

In the present work, we adopt Scheme 2, which while is more difficult to program, but more accurate and reliable as shown in the present work. In Scheme 1, any inaccurate DFT predictions at a given volume (as well as temperature) result in inaccurate $F(V,T)$ for all $T$. Using Scheme 2, we can use fewer data points for time-consuming calculations (e.g., a minimum of two volume points for phonon calculations) and remove inaccurate or failed data points during each fitting. Furthermore, different models and model parameters can be selected to fit $E$, $F_{vib}$, and $F_{el}$ individually and give a more accurate $F(V,T)$ curve.

Based on our previous study of equation of state (EOS) [3,6], the linear Birch-Murnaghan (BM) equation [3] is selected as the present model to fit $E$, $F_{vib}$, and $F_{el}$ as a function of $V$,

$$f(V) = a_1 + a_2 V^{-2/3} + a_3 V^{-4/3} + a_4 V^{-6/3} + a_5 V^{-8/3} \qquad \text{Eq. 2}$$

where $a_1$ to $a_5$ are model/fitting parameters. For $E(V)$ at 0 K, Eq. 2 is the $E$-$V$ EOS and can be used to estimate properties at a given external pressure, including equilibrium volume $V_0$, total energy $E_0$, bulk modulus $B_0$ and its first and second pressure derivatives of $B' = dB/dP$ and $B'' = d^2B/dP^2$ [3]. Note that at least 3 parameters ($a_1$, $a_2$, and $a_3$) are required to determine $V_0$. By adding the 4[th] parameter $a_4$, $B'$ can be estimated, which is a key indicator to evaluate thermal expansion. For example, $B' = 0$ corresponds to zero thermal expansion and a larger $B'$ corresponds to a larger thermal expansion [3]. Similarly, by adding the 5[th] parameter $a_5$, $B''$ can be determined. However, more parameters (e.g., with $a_5$) may result in overfitting of the $E$-$V$ curve. Besides Eq. 2, more linear and nonlinear models (i.e., the EOS's) have been previously reported [3].

To determine the optimal number of model/fitting parameters in Eq. 2, we employ the Akaike information criterion (AIC) [7] by estimating the quality of a collection of models and model parameters for a large dataset as follows,

$$\text{AIC} = 2k + n\log(\text{RSS}) \text{ with RSS} = \frac{\sum_{i=1}^{n}\left(y_i^{model}-y_i\right)^2}{n} \qquad \text{Eq. 3}$$

where $k$ and $n$ are the numbers of model parameters and data points, $y_i^{model}$ and $y_i$ are the fitted model result and the true result for data point $i$, respectively. For small datasets such as the DFT-based results of $E$, $F_{vib}$, and $F_{el}$, a corrected AIC (i.e., the AICc [8,9]) is adopted,

$$\text{AICc} = \text{AIC} + \frac{2k^2+2k}{n-k-1} \text{ when } n > k+1 \qquad \text{Eq. 4}$$

$$\text{AICc} = \text{AIC} + (2k^2+2k)(-n+k+3) \text{ when } n \leq k+1. \qquad \text{Eq. 5}$$

Eq. 5 is a correction suggested by Bocklund [8] for the case of $n \leq k+1$.

Following our previous efforts using DFT-based QHA (see [2,3] and their citations), we revisit Eq. 1 using the above methodology through predicting the instantaneous, linear coefficient of thermal expansion (CTE or $\alpha_L$) of IN625 at ambient pressure (herein, $P = 0$). The linear CTE is defined as,

$$\alpha_L = \left(\frac{d\varepsilon_0^T}{dT}\right)_P = \frac{1}{3V_0^T}\left(\frac{dV_0^T}{dT}\right)_P \qquad \text{Eq. 6}$$

where $\varepsilon_0^T$ is thermal strain at $T$, measured in the present work as detailed in Supplementary material, and $V_0^T$ is equilibrium volume at $T$ determined by Eq. 1.

For the face-centered cubic (FCC) phase in IN625 [10], we employ the relatively small supercells by considering both computational efficiency and different local environments of the concentrated alloy. The composition is approximated as $Ni_{21}Cr_8Mo_2Nb_1$, which is close to the major composition of IN625, i.e., Ni - 20.61Cr - 8.82Mo - 3.97Nb (in wt. %) [11], see details in the Supplementary Excel file. Here, we use two methods to generate diverse supercells, i.e., six 32-atom special quasirandom structures (SQSs) [12] by ATAT [13] and six 32-atom supercells in random approximates (SCRAPs) [14]. All these 12 supercells were initialized in cubic symmetry. One of the SCRAPs (marked as SCRAP$_f$) and its structure file are provided in Supplementary Figure S 1 and Supplementary Table S 1. The selected, representative SCRAP$_f$ is considered due to its smooth $E$-$V$ data points and the resulted equilibrium properties in the middle of these supercells as shown in Supplementary Table S 2. All DFT-based total energy and phonon calculations were performed by VASP [15] with details in Supplementary Material.

Figure 1 shows the fitted *E-V* curves using the 4- and 5-parameter BM EOS's (see Eq. 2, labeled by BM4 and BM5, respectively), superimposed with the data points from configuration SCRAP$_f$ by DFT-based calculations. Both fitted curves with BM4 and BM5 are almost identical to each other. However, the AICc value by BM4 fitting is much lower than that by BM5 (-29.7 vs. 9.0), indicating the overfitting trend by BM5. Equilibrium properties determined by BM4 for the random 12 configurations of IN625 are listed in Table S 2, showing that, $E_0 = -7.007 \pm 0.009$ (eV/atom), $V_0 = 11.691 \pm 0.066$ (Å$^3$/atom), $B_0 = 187.7 \pm 10.9$ (GPa), and $B' = 5.95 \pm 0.78$; where the values after the symbol $\pm$ indicate the standard deviations. We note that the $B'$ value of IN625 is larger than those of the pure elements with FCC lattice of which the alloy is comprised, i.e., 4.2 for Cr, 4.9 for Ni, 3.7 for Nb, and 4.1 for Mo (see details in Table S 3). The increased $B'$ value indicates a larger CTE of IN625 than those of the pure elements, which would result in a less accurate CTE estimation if using a rule-of-mixture method based on CTE values of pure elements for IN625 as shown in Figure S 2.

Figure 2a and b show the fitted vibrational contribution $F_{vib}(V,T)$ via phonon DOS's with one example plotted in Figure S 3 and thermal electronic contribution $F_{el}(V,T)$ via electronic DOS's with one example plotted in Figure S 4, for the configuration SCRAP$_f$ at 1000 K, together with DFT-based predictions superimposed. The vibrational contribution was calculated using phonon DOS as opposed to the Debye model for enhanced accuracy and thermal electronic contribution was calculated following the Fermi-Dirac distribution. The AICc values, plotted in Figure S 5 for various model fits to the DFT-predicted energy contributions, indicate that the 3-parameter BM equation (BM3) is the best to fit $F_{vib}(V,T)$, while BM2 is the best to fit $F_{el}(V,T)$ at 1000 K since the $F_{el}(V,T)$ curve by DFT are in general scattered due to the non-smooth characteristic of electronic DOS near the Fermi level as shown in Figure S 4.

AICc analysis suggests fitting the DFT results with 4, 3, and 2 model terms, i.e., BM4, BM3, and BM2 in Eq. 2, for $E$, $F_{vib}$, and $F_{el}$ as a function of volume, respectively. With these models, Helmholtz energy $F(V,T)$ as well as the equilibrium properties in terms of its derivatives such as $V_0$, $F_0$, $B_0$, and $B'$ at each $T$ can be predicted for IN625 at external pressure $P = 0$ GPa.

Figure 3 plots the predicted $V_0^T$ and linear CTE (via Eq. 6) of IN625 based on DFT-based QHA for SCRAP$_f$. Figure 3a shows that the predicted $V_0^T$ agrees well with experimental results estimated by the present thermal strain data with the raw data presented in Supplementary Excel file. Here we assume that the $V_0^T$ values at 298 K are equal for both the predicted and the measured results, since the measured $V_0^{298}$ was not reported and it is not necessary for thermal strains. As a contrast, using the BM2 to fit $F_{vib}(V,T)$ results in an apparent difference between the predicted and experimental $V_0^T$. For example, the difference is up to 0.046 Å$^3$/atom at 1000 K. Figure 3b shows that the presently predicted CTE values (the red line) are in good agreement with the present measurements (except for the tail part up to 450 K) and those in the literature by Heugenhauser et al. [16]. Note that the lower tail of the measured CTE may be unrealistic due to the dilatometer used. These data (from room temperature to 450 K) are hence plotted for reference only.

As shown in Table S 2, DFT-based equilibrium properties at 0 K possess uncertainties due to the relatively small supercells used. Figure 4 depicts the sensitivity of CTE due to different equilibrium properties at 0 K, i.e., $V_0$, $B_0$, and $B'$, through changing only the $E(V)$ term in Eq. 1 and remaining unchanged for $F_{vib}(V,T)$ and $F_{el}(V,T)$ which are the same as those for SCRAP$_f$. Within the ranges of the standard deviations for equilibrium properties of IN625 as shown in Table S 2, Figure 4 shows that the linear CTE increases greatly by increasing $B'$ or by decreasing $B_0$, while it increases only slightly by decreasing $V_0$. For example, the linear CTE value increases about 2×10$^{-6}$ K$^{-1}$ by increasing $B'$ by 0.8 or by decreasing 10 GPa of $B_0$. However, decreasing $V_0$ by 0.07 Å$^3$/atom results in a slight increase of linear CTE by 0.05×10$^{-6}$ K$^{-1}$. It is worth mentioning that the fitted $B'$ value is very sensitive to the quality of DFT-based calculations, following by $B_0$ and then the less sensitive $V_0$. For example, Table S 2 shows that the changes are 0.5 for $B'$ (5.68 vs. 6.18), 2.5 GPa for $B_0$ (182.5 vs. 179.9), and 0.001 Å$^3$/atom for $V_0$ (11.703 vs. 11.702) based on two BM4 fittings using the relaxed data points and the static DFT calculations for SCRAP$_f$, respectively. We hence conclude that $B'$ is a dominant contribution, greatly influencing the predicted CTE and other thermodynamic properties of IN625.

To further analyze uncertainty of the predictions, we initially assume that the predicted $V_0$, $B_0$, and $B'$ are normally distributed and use the predicted results in Table S 2 to build their normal distribution functions. From these distributions, we randomly select 200 datasets of $V_0$, $B_0$, and $B'$.

Figure S 7 shows that the selected datasets are approximately normally distributed for each property. Using these 200 datasets and the same methodology as used to predict Figure 4, 200 CTE curves are generated as a function of temperature as plotted in Figure S 8 at 1000 K, together with the fitted Weibull distribution and two *highest density intervals* (HDIs) of the Weibull distribution, i.e., the 40% HDI and the 80% HDI. It shows that the predicted CTE values do not follow normal distribution. Instead, the two-parameter Weibull distribution function can be used as a more appropriate description of the CTE distribution. Figure 3b plots these two HDIs as a function of temperature, showing that the red line from DFT-based QHA for SCRAP$_f$ is at the top of the 40% HDI area, and most of experimental results are at the top of 80% HDI area. In fact, the CTE predictions due to the uncertainties of $V_0$, $B_0$, and $B'$ are in the region with higher CTE values as shown in Figure S 8.

In summary, we revisit the DFT-based QHA by examining the best model selection for each physical contribution, i.e., the static total energy $E$, phonon $F_{vib}$, or thermal electron $F_{el}$, to free energy in terms of the Akaike information criterion with a correction (AICc). In general, the suggested model parameters are 4, 3, and 2 in Eq. 2 to fit $E$, $F_{vib}$, and $F_{el}$ as a function of volume, respectively. Taking the concentrated Inconel alloy 625 (IN625) as an example, we find that the DFT-based QHA can accurately predict thermodynamic properties by optimizing the model fitting of each physical contribution. For example, the predicted linear coefficient of thermal expansion (CTE) agrees well with the present measurements and those in the literature. In addition, the present work indicates that the predicted CTE as well as other thermodynamic properties is sensitive to equilibrium properties at 0 K from EOS fitting. The dominant contribution to the properties is the pressure derivative of bulk modulus ($B'$), followed by bulk modulus ($B_0$) and the less sensitive volume ($V_0$). Using various random configurations to represent different local environments for a given alloy, it is possible to examine uncertainty of the predicted properties using different equilibrium properties determined. For the present CTE case, we use the normal distributions to generate probability sampling of $V_0$, $B_0$, and $B'$, and the Weibull distribution to fit the predicted CTE values; showing that the *highest density interval* (HDI) is at the top of the predicted CTE values for IN625.

**Acknowledgements**

This work was funded by NIST Award 70NANB22H051 and by the eXtremeMAT (XMAT) project of U.S. Department of Energy (DOE) and the DOE High Performance Computing for Energy Innovation (HPC4EI), and through an appointment to DOE Faculty Research Program at the National Energy Technology Laboratory (NETL) administered by the Oak Ridge Institute for Science and Education (S. L. Shang). First-principles calculations were performed partially on the Roar supercomputers at the Pennsylvania State University's Institute for Computational and Data Sciences (ICDS), partially on the resources of the National Energy Research Scientific Computing Center (NERSC) supported by the DOE Office of Science User Facility operated under Contract no. DE-AC02-05CH11231 using NERSC award ALCC‑ERCAP0022624, and partially on the resources of the Extreme Science and Engineering Discovery Environment (XSEDE) supported by NSF with Grant no. ACI-1548562.

**Figures and Figure Captions**

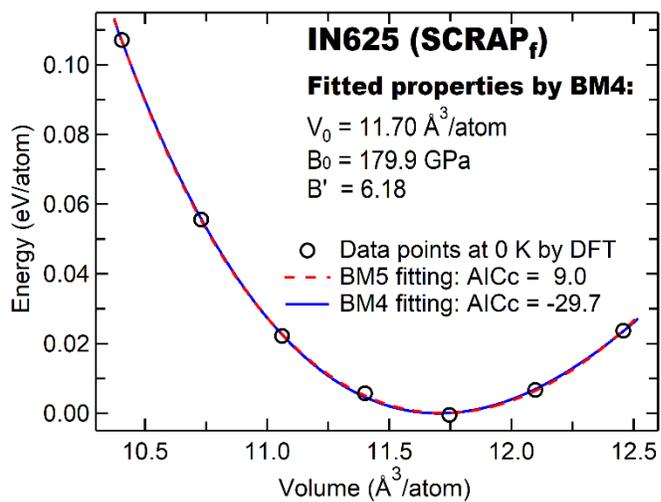

Figure 1. Data points at 0 K by DFT-based final static calculations for the configuration SCRAP$_f$, together with the fitted *E-V* EOS curves using 4 parameters (BM4) and 5 parameters (BM5) of Eq. 2. The fitted properties (see Table S 2) and the calculated AICc values are also listed on this plot.

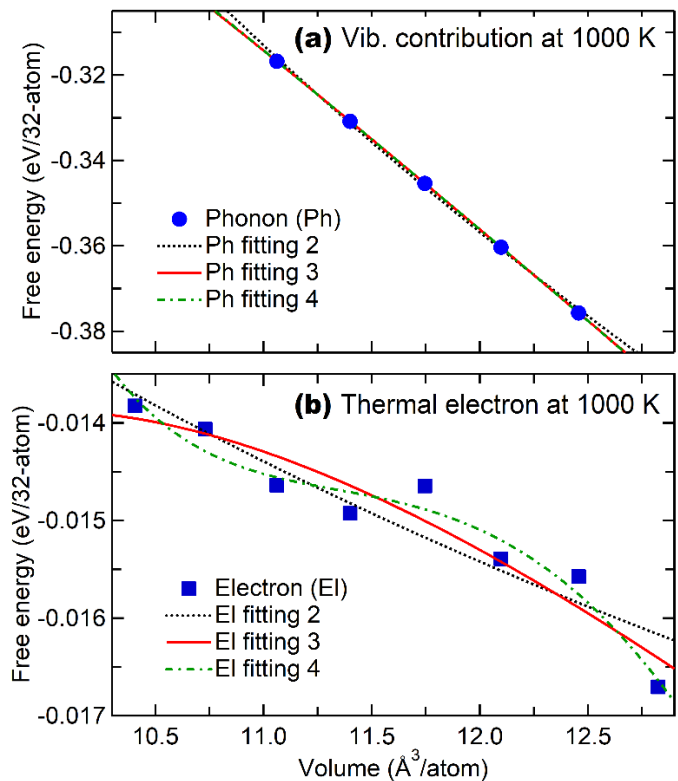

Figure 2. Calculated and fitted free energy for (a) $F_{vib}(V,T)$ and (b) $F_{el}(V,T)$ at 1000 K for the configuration SCRAP$_f$. The lowest AICc values correspond to the BM3 fitting for phonon ($F_{vib}(V,T)$) and the BM2 for thermal electron ($F_{el}(V,T)$) using Eq. 2 with more details in Figure S 5.

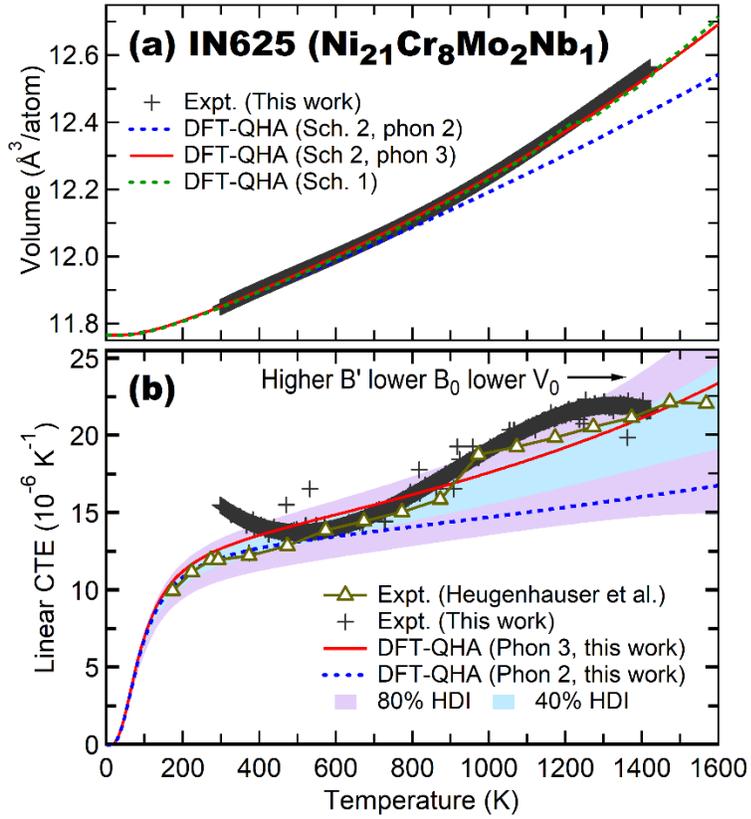

Figure 3. Predicted (a) volume and (b) linear CTE of IN625 by DFT-based QHA (by Scheme 2 (Sch. 2) with $F_{vib}$ fitted by 2 or 3 parameters using Eq. 2 at each $T$, labeled as phon 2 and phon 3, respectively) in comparison with the present experiments with raw data in Supplementary Excel file and those by Heugenhauser et al. [16]. The shaded areas indicate the percentages of *highest density interval* (HDI) calculated using the Weibull distribution (*cf.*, Figure S 8). The predicted CTE by QHA Scheme 1 (Sch. 1) is shown in Figure S 6 for comparison.

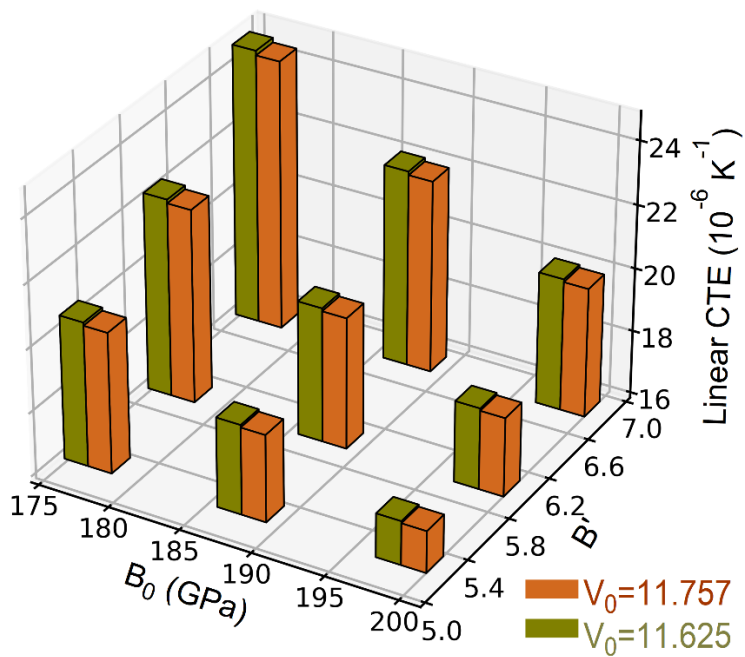

Figure 4. Sensitivity of the predicted linear CTE of IN625 at 1500 K with respect to the changes of equilibrium properties at 0 K with data in Table S 2, i.e., $B_0$ (187.7 ± 10.9 GPa), $B'$ (5.95 ± 0.78), and $V_0$ (11.691 ± 0.066 Å$^3$/atom).

Supplementary Material for:

# Revisiting first-principles thermodynamics by quasiharmonic approach:
# Application to study thermal expansion of additively-manufactured Inconel 625

Shun-Li Shang,[1,2,*] Rushi Gong,[1] Michael C. Gao,[2] Darren C. Pagan,[1] and Zi-Kui Liu[1]

1. Department of Materials Science and Engineering, The Pennsylvania State University, University Park, PA 16802, USA
2. Computational Science Engineering Directorate, National Energy Technology Laboratory, Albany, OR 97321, USA

*E-mail: sus26@psu.edu (S. L. Shang)

**Experimental details to measure thermal expansion**

Detailed methodology to prepare the additively-manufactured IN625 alloy was reported at the NIST website [1].

For thermal expansion of In625, a dilatometry specimen was extracted using electrical discharge machining (EDM) to have a nominal 10 mm length and 4 mm diameter. EDM scale was removed from the sample by progressively grinding with silicon-carbide papers (P180, P360, and P500). An S-type thermocouple with a precision of 0.05°C for use with a DIL 805 A/D (TA Instruments, New Castle, DE) was then spot welded with current and time settings of 5 and 4, respectively, under a nitrogen cover gas. The sample was mounted in a DIL 805 A/D dilatometer with an alpha sled installed which has a precision of 0.05 µm. Good contact between the sample ends and the pushrods was obtained by gently rotating the specimen during mounting. The linear variable differential transformer (LVDT) displacement was manually set to zero to keep the test in the high resolution (20 mV/µm) displacement range. A vacuum of $2\times10^{-5}$ mbar was pulled to remove oxygen from the system to minimize sample oxidation which could restrict specimen expansion [2]. The first temperature profile started at room temperature and increased to 1473 K at a rate of 0.08 K/min. Quenching to room temperature was performed using the high purity compressed helium (UN1046) cooled by liquid nitrogen (UN1977, Linde, Danbury, CT) to achieve a consistent rate of approximately to 100 K/min. Alumina pushrods with a thermal expansion of $7.4\times10^{-6}$ $K^{-1}$ [3] was used as they would not soften at this temperature. A follow-on experiment was performed with a maximum temperature of 1423 K such that the fused silica pushrods with a thermal expansion of $0.4\times10^{-6}$ $K^{-1}$ [3] could be used. For this test, the heating and cooling cycles were repeated three times. Both tests were corrected to account for thermal expansion of the system by running a Pt-reference sample through the same thermal profiles. Correction to the data was then applied using the DIL805 software module. Sample temperature, displacement, and the first and second derivatives of temperature with respect to time for each time step are output from the test. Note that this procedure follows the standard convention for linear thermal expansion measurement [4].

**DFT-based first principles calculations**

All DFT-based first-principles calculations were performed using the VASP code [5]. The ion-electron interaction was described by the projector augmented wave (PAW) method [6] and the exchange-correlation functional was depicted by the generalized gradient approximation (GGA) developed by Perdew, Burke, and Ernzerhof (PBE) [7]. In VASP calculations, electron configurations for each element were the same as those used by the Materials Project [8]; the automatic $k$-point meshes were adopted with a value of 25 to determine $k$-point meshes for the 32-atom supercells (i.e., the six SQS's [9] and six SCRAPs [10]); and the VASP setting of "PREC = Accurate" was used to determine the plane-wave basis set for both structural relaxations and phonon calculations. The energy convergence criterion of the electronic self-consistency was at least $10^{-6}$ eV/atom for all calculations. The reciprocal-space energy integration was performed by the Methfessel-Paxton technique [11] for structural relaxations. For the selected supercell of SCRAP$_f$, the final static calculations of total energies and the electronic density of states (DOS's) were calculated by the tetrahedron method with a Blöchl correction [12] using a wave cutoff energy of 520 eV.

Phonon calculations were performed for SCRAP$_f$ using the supercell approach [13] in terms of the YPHON code [14]. Here, the VASP code was again the computational engine to calculate force constants using the finite differences method. The employed supercell, the corresponding $k$-point meshes, and the other settings are the same as the aforementioned structural relaxations. Due to magnetic natures of the elements Ni and Cr, DFT-based calculations were performed using the spin polarization calculations in terms of the ferromagnetic configurations.

# Three Supplementary Tables

Table S 1. Relaxed structure/configuration in terms of the VASP POSCAR format [15] for one of the SCRAPs, i.e., the SCRAP$_f$ in Table S 2, see also Figure S 2.

```
Relaxed structure for the 32-atom configuration of SCRAPf
1.0000000000
   7.2301742366      0.0310644386      0.0056118570
   0.0315390740      6.9938207888     -0.0067402691
   0.0056085333     -0.0065488351      7.2143420441
Ni  Cr  Mo  Nb
21   8   2   1
Direct
  0.9992407335   0.2525256106   0.2464628932 Ni
  0.0022901634   0.2500248922   0.7514006262 Ni
  0.9990231168   0.7512303070   0.7533659897 Ni
  0.4992138304   0.7530195417   0.7453578102 Ni
  0.2363896856   0.0016636840   0.2481802193 Ni
  0.2432966877   0.4962735917   0.2467386422 Ni
  0.7586272583   0.4963419029   0.2497490769 Ni
  0.2468194078   0.9970916786   0.7511652240 Ni
  0.7570458595   0.0045663465   0.7524725684 Ni
  0.2406611427   0.5018090729   0.7537023683 Ni
  0.7607465501   0.4994007609   0.7508519801 Ni
  0.2443656566   0.2488194038   0.9954826681 Ni
  0.7569353870   0.2527851382   0.9952893533 Ni
  0.2398042592   0.7509285889   0.9964214588 Ni
  0.7563242394   0.7489224831   0.9967968436 Ni
  0.2449072148   0.2454594140   0.5036836401 Ni
  0.2457733069   0.7524180127   0.5041838944 Ni
  0.5000367682   0.9978608183   0.9881433584 Ni
  0.4985459876   0.4982676752   0.9862317001 Ni
  0.0008177734   0.9989844420   0.5035080500 Ni
  0.0000000000   0.5042122567   0.4989548776 Ni
  0.9963137413   0.7530976702   0.2516837127 Cr
  0.4992639518   0.2413172038   0.7490041645 Cr
  0.7679833464   0.0052265354   0.2496622172 Cr
  0.7706131581   0.2492944888   0.5039584991 Cr
  0.7596983703   0.7596721502   0.5161631715 Cr
  0.9985075254   0.0018109145   0.9976231974 Cr
  0.9987358115   0.5024710066   0.0026388836 Cr
  0.4966721444   0.0096743847   0.5162112296 Cr
  0.5014508976   0.2384090715   0.2356176733 Mo
  0.4972710925   0.4857842434   0.5163661274 Mo
  0.4972655792   0.7506367092   0.2429278810 Nb
```

Table S 2. Predicted equilibrium properties of IN625 (i.e., the FCC-based $Ni_{21}Cr_8Mo_2Nb_1$) by DFT-based first-principles calculations without the final static calculations except for the marked one of $SCRAP_f$.

| Configurations | $E_0$ (eV/atom) | $V_0$ (Å³/atom) | $B_0$ (GPa) | $B'$ |
|---|---|---|---|---|
| $SCRAP_a$ | -7.0079 | 11.682 | 194.3 | 5.25 |
| $SCRAP_b$ | -7.0121 | 11.648 | 195.0 | 5.78 |
| $SCRAP_c$ | -6.9909 | 11.736 | 181.1 | 5.72 |
| $SCRAP_d$ | -7.0007 | 11.624 | 206.9 | 5.12 |
| $SCRAP_e$ | -7.0165 | 11.697 | 180.4 | 6.27 |
| $SCRAP_f$ | -7.0236 | 11.703 | 182.5 | 5.68 |
|  |  | 11.702* | 179.9* | 6.18* |
| $SQS_a$ | -7.0058 | 11.643 | 182.8 | 6.86 |
| $SQS_b$ | -7.0045 | 11.823 | 176.2 | 4.70 |
| $SQS_c$ | -7.0175 | 11.679 | 204.6 | 5.95 |
| $SQS_d$ | -7.0003 | 11.638 | 187.6 | 7.49 |
| $SQS_e$ | -7.0003 | 11.622 | 190.5 | 6.66 |
| $SQS_f$ | -7.0047 | 11.800 | 171.0 | 5.97 |
| AV ‡ → | -7.007 | 11.691 | 187.7 | 5.95 |
| SD ‡ → | 0.009 | 0.066 | 10.9 | 0.78 |

* Results after the final static calculations, *cf.*, Figure 1.
‡ AV indicates the average result and SD the standard deviation. There values are also reported in Table S 2.

Table S 3. Equilibrium properties of pure elements in FCC lattice by DFT-based calculations using different exchange correlation (X-C) functionals of LDA, GGA-PBE, GGA-PBEsol, and GGA-PW91 [16], where "av" indicates the average results.

| Elements | X-C functionals | $V_0$ (Å³/atom) | $B_0$ (GPa) | $B_0$-av | $B'$ | $B'$-av |
|---|---|---|---|---|---|---|
| Cr | LDA | 11.122 | 276.6 | 252.6 | 4.15 | 4.2 |
|  | PBE | 11.857 | 237.3 |  | 4.17 |  |
|  | PBEsol | 11.428 | 259.6 |  | 4.11 |  |
|  | PW91 | 11.901 | 236.7 |  | 4.18 |  |
| Ni | LDA | 10.022 | 252.3 | 217.4 | 4.90 | 4.9 |
|  | PBE | 10.882 | 195.3 |  | 4.88 |  |
|  | PBEsol | 10.373 | 226.6 |  | 4.87 |  |
|  | PW91 | 10.920 | 195.6 |  | 4.90 |  |
| Nb | LDA | 17.987 | 185.0 | 173.4 | 3.74 | 3.7 |
|  | PBE | 18.967 | 165.7 |  | 3.70 |  |
|  | PBEsol | 18.345 | 177.3 |  | 3.73 |  |
|  | PW91 | 18.975 | 165.7 |  | 3.79 |  |
| Mo | LDA | 15.277 | 268.9 | 251.5 | 3.96 | 4.1 |
|  | PBE | 16.019 | 239.5 |  | 4.06 |  |
|  | PBEsol | 15.528 | 258.4 |  | 3.96 |  |
|  | PW91 | 16.195 | 239.4 |  | 4.21 |  |

**Seven Supplementary Figures**

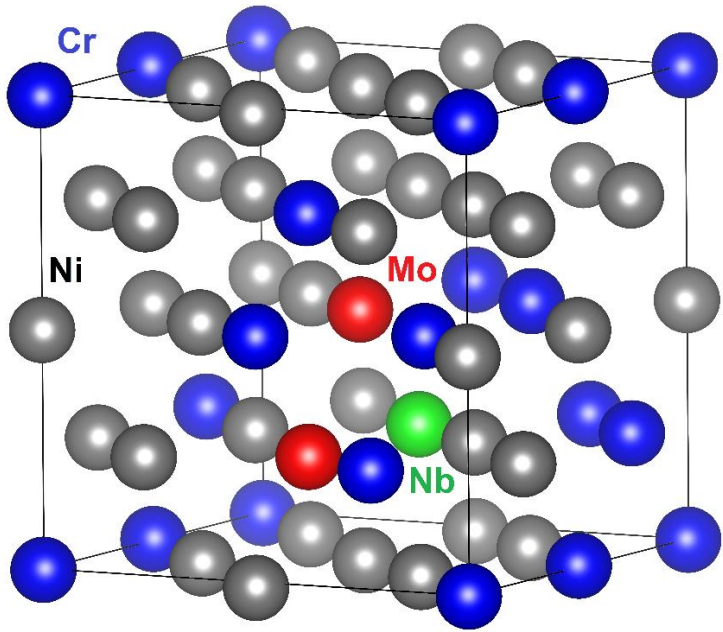

Figure S 1. One of the 32-atom supercells for IN625 (i.e., the configuration SCRAP$_f$) based on the composition of Ni$_{21}$Cr$_8$Mo$_2$Nb$_1$.

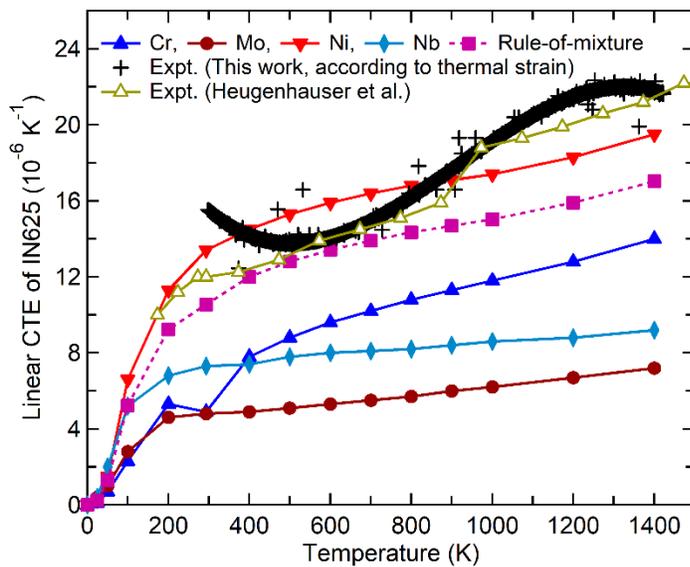

Figure S 2. Experimental linear CTE values of pure elements as recommended by Touloukian et al. [17]. The results of IN625 are based on the present measurements and those by Heugenhauser et al. [18]. The predicted CTE values of FCC-based Ni$_{21}$Cr$_8$Mo$_2$Nb$_1$ are based on the rule-of-mixture of CTE values of pure elements (in this figure).

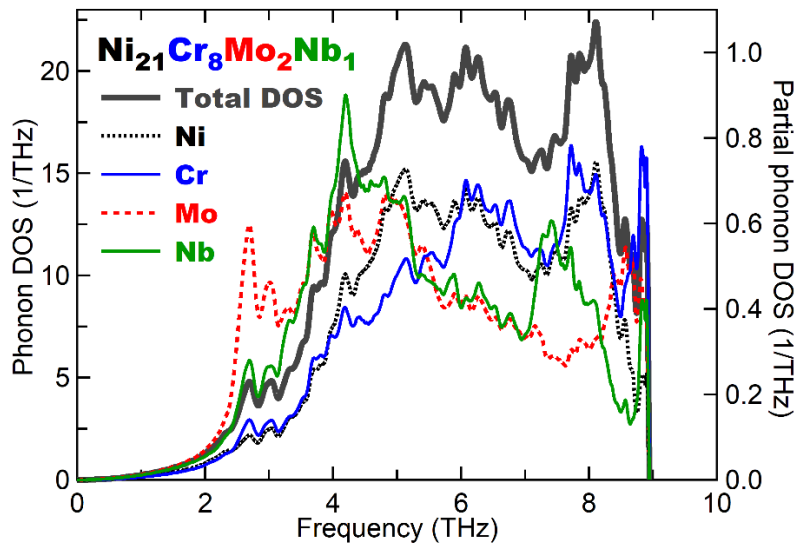

Figure S 3. Total phonon density of states (DOS) and partial phonon DOS per atom of the element Ni, Cr, Mo, or Nb for $Ni_{21}Cr_8Mo_2Nb_1$ at the equilibrium volume of $SCRAP_f$. It shows that the configuration $SCRAP_f$ is stable without imaginary phonon modes, and the element Mo has relatively large contribution to free energy since its phonon DOS possesses higher density at the low frequency region (e.g., < 3 THz), following by Nb, then Cr and Ni [19–21].

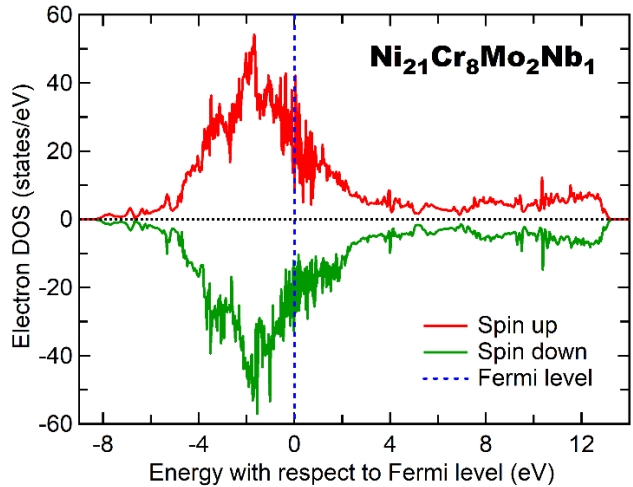

Figure S 4. Electronic density of states of $Ni_{21}Cr_8Mo_2Nb_1$ at the equilibrium volume of $SCRAP_f$.

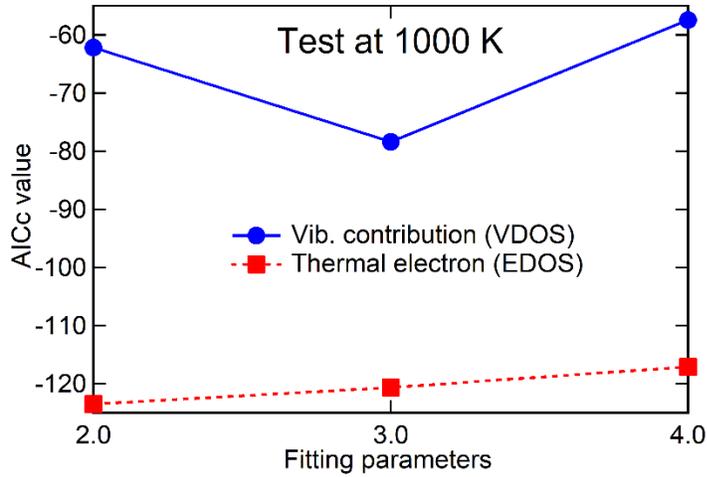

Figure S 5. Predicted AICc values regarding vibrational contribution $F_{vib}(V,T)$ via phonon DOS's and thermal electronic contribution $F_{el}(V,T)$ via electronic DOS's to free energy at 1000 K; see Figure 2.

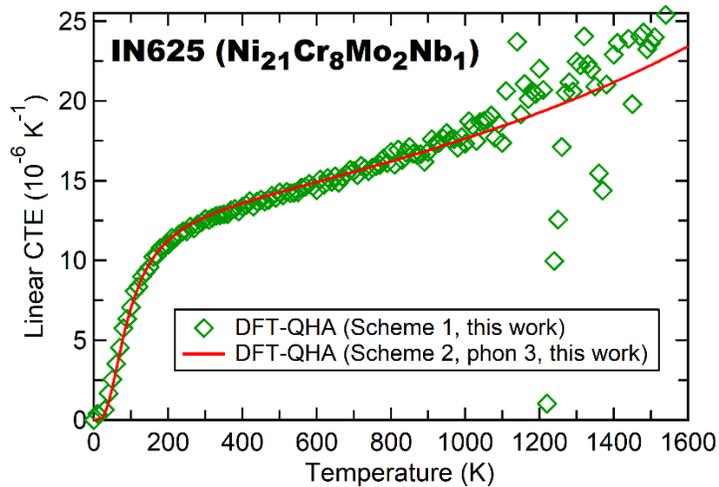

Figure S 6. Predicted values of linear CTE of IN625 by DFT-based QHA using Scheme 1 (Sch. 1) and Scheme 2 (Sch. 2, with $F_{vib}$ fitted by 3 parameters using Eq. (2) at each $T$, labeled as phon 3), respectively, see also Figure 3b for more CTE results.

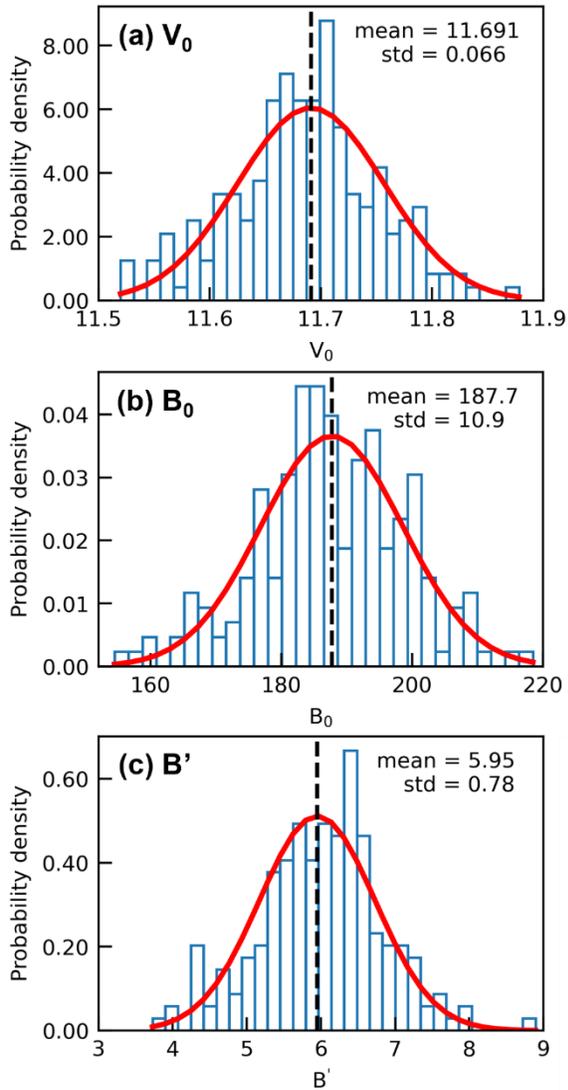

Figure S 7. Distributions of the randomly selected 200 datasets of $V_0$ (a), $B_0$ (b), and $B'$ (c) by assuming each of them normally distributed in terms of the predicted properties in Table S 2 (also shown in the figures). It shows that the presently selected 200 datasets (plotted as bars) follow roughly the normal distributions as indicated by the red lines.

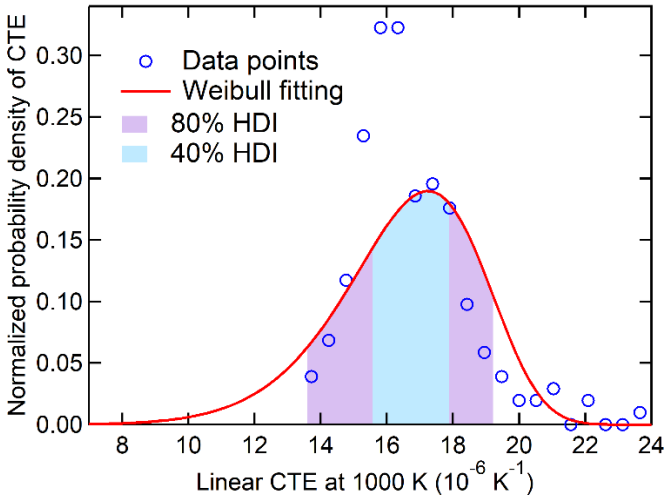

Figure S 8. Distribution of the predicted CTE values of IN625 at 1000 K according to 200 predictions using the selected 200 datasets in Figure S 7. The red line shows the fitting of Weibull distribution, and two areas show the percentages of *highest density interval* (HDI) according to the Weibull distribution.

**One Supplementary Excel File (Ask Dr. Shang for this file)**

**Sheet of "625-composition":**
This sheet shows the compositions of IN625 based on the benchmark tests at NIST [1], in comparison with the compositions $Ni_{21}Cr_8Mo_2Nb$ used in the present 32-atom supercells.

**Sheet of "Raw-Data":**
This sheet shows (i) the raw data of measured CTE data of pure elements (Cr, Mo, Ni, and Nb) as recommended by Touloukian et al. [17], and (ii) the present measured thermal strains and the estimated CTE results.